\documentclass[12pt]{article}
\usepackage{amsmath}
\usepackage{amssymb}
\renewcommand{\title}[1]{%
    \bigskip%
    \begin{center}%
    \Large\bf #1%
    \end{center}%
    \vskip .2in}

\renewcommand{\author}[1]{%
   {\begin{center}
    #1
    \end{center}}}
\newcommand{\address}[1]{\vspace{-1.7em}\vspace{0pt}
    {\begin{center}
    \it #1
    \end{center}}}

\begin{document}
\title{\bf{ Geometry of  Nonrelativistic string}}

\author
{  
Sk. Moinuddin   $\,^{\rm a, b}$, 
Pradip Mukherjee  $\,^{\rm a,c}$}

\address{$^{\rm a}$Department of Physics, Barasat Government College,Barasat, India}
\address{$^{\rm b}$\tt
dantary95@gmail.com  }
\address{$^{\rm c}$\tt mukhpradip@gmail.com}

\abstract{ The nonrelativistic bosonic string theory in a curved  manifold is formulated here  using gauging of symmetry approach ( Galilean Gauge theory ) .  The corresponding model in flat space has some global symmetries . By localizing these symmetries as per Galilean Gauge theory , the action for the nonrelativistic string interacting with gravity is obtained. A canonical analysis of the model has been performed which demonstrate that the transformations of   the basic field  variables under gauge transformations in phase space are equivalent to the  diffeomorphism parameters by an exact mapping. Thus complete consistency of our results from both  Lagrangian and Hamiltonian procedures are established.

}

\vskip 2cm

  \section{Introduction}
The essence of Einstein's general theory of relativity is the assumption that gravity  generated by massive bodies is not an interaction force but curvature of the spacetime manifold . Motion of the  test objects take place in this curved spacetime manifold produced . It is always assumed that the test object does not influence the existing field \cite{MTW}. Then the motion of an otherwise free  particle is a solution of the geodesic equation. If gravity is not there, then geometry of spacetime becomes the Minkowski manifold with zero curvature.  The General relativity (GR) is thus compatible  with the special theory of relativity (STR). Newtonian gravity is not. The  Newton's theory of universal  gravitation implies that gravitational force is transmitted instantaneously, thereby violating the second postulate of STR. But interestingly the dynamics of Newtonian gravity also satisfies the principle of equivalence. Building on this, Cartan formulated Newtonian gravity as a geometric theory \cite{Cartan-1923}\cite{Cartan-1924} in Newton-Cartan (N-C) spacetime. Note that the test objects which move in the field (like particles , strings)  are supposed not to disturb the field configuration. This last point is shared by all field theories. 
 
 \hskip.2cm  Diffeomorphism invariance ($x^\mu\to x^\mu+\xi^{\mu}$) in a spacetime manifold is called nonrelativistic  diffeomorphism invariance (NRDI) if in the locally inertial system at any spacetime point , time and space are considered on different footing. Further if the direction of time flow is consider absolute , then the spacetime becomes degenerate that is metric properties can not be defined with a single metric . Indeed, one can demonstrate the transition by examining the structure of the metric in the $c \to \infty$ limit to pass into a collection of singular metrics  of rank 1 and 3 respectively \cite{ABPR}. Different  nonrelativistic geometries may follow from the limiting procedure \cite{BGM} .Thus , contrary to what appears in the first sight to be  very easy, taking the nonrelativistic limit has many subtle issues  \cite{BGM1}. There lies the importance of the  nonrelativistic diffeomorphism which provides directly the coupled theory with gravity.

 \hskip.2cm The nonrelativistic diffeomorphism has become a topic
  of current interest due to its applications in such frontal 
  areas as quantum gravity \cite {Ho} , fractional quantum Hall effect \cite{SW} \cite{SW1} and other new branches of condensed matter physics and many others. Thus the direct coupling of a nonrelativistic theory with gravity ( i.e; formulation of nonrelativistic diffeomorhism invariance) becomes important.  
  
 \hskip.2cm In this connection it may be mentioned that we have proposed a method of generating NRDI. The algorithm is as follows:  Consider a theory invariant under the Galileon transformations in the Euclidean space. An interesting   correspondence is unveiled by localizing symmetry transformation parameters  under the Galileon group of transformations in the flat limit of Newton-Cartan space. Fields would transform formally in the same way but the derivatives transform differently as the parameters now are no more constants. The symmetry is regained when appropriate gauge fields are included which transform properly. These transformation of the fields have one to one correspondence with the transformations of the vielbeins and spin connections of the Nerwton - Cartan manifold. The method is inspired by the well known Poincare Gauge theory \cite{Uti} and hence named as Galileon Gauge theory (GGT) \cite{BM4}. This theory was proposed in \cite{BMM1} and \cite{BMM2}  to addressed a paradox arrising from Son and Wingate's \cite{SW} seminal paper. The formal structure of GGT was subsequently extended to include the gauge fields successfully \cite{BMM3}. Meanwhile, simultaneously the theory was applied to explain  the interaction of  electron in the Chern Simon model \cite{frad} and the nonrelativistic nature of the Horava-Lifshitz geometry \cite{Ho}. Thus in this way a compact theory of achieveing NRDI was developed which was christened as Galileon Gauge theory, for reasons we have just discussed.  Hence forth the GGT has been applied to numerous problems in the domain  of the nonrelativistic theories including the motion of particles (with and without spin), fields etc in the nonrelativistic backdrop \cite{BMN1}-\cite{Bn} . {\bf {However one important system that is string or higher branes is yet to be treated by GGT. So this is one motivation to under take the present project.}}
 
 \hskip .5cm {\bf{ The motivation just mentioned was enhanced hundredfolds by the results from the nonrelativistic string \cite{bag1}\cite{bag}  ,the essence of which is the inadequacy of Newton-Cartan  spacetime in describing the string motion.}} There conclusion was to generalized  the Newton-Cartan algebra so that the stringy motion can be described. However note that the standard Newton-Cartan algebra was found to be appropriate for nonrelativistic particle etc. So the  stringy extension should go to the standard Newton-Cartan by compactifing the string. Such an analysis is yet to be done. Meanwhile it is really a singular fact that the test object (here string) changing the geometry. {\bf{Naturally one would be eager to know what happens when we use the GGT algorithm. In this paper we have also address this point.}}

%Gravitation is the weakest of the fundamental forces of nature. Notwithstanding this, the physics in the large scale is overwhelmingly influenced  by the gravitational interaction. The reason is,  gravitation is always attractive unlike the other field theories . Einstein in his celebrated General (Theory of ) relativity (GR) formulated the gravitational  interaction as the change of curvature of space-time . Einstein's
%equations  determine the curvature produced by a certain mass energy distribution. Massive object which are not interacted by any other forces, follow ``straight lines"\footnote{Technically called the geodesic.} in this curved manifold. By assumption the test objects  do not disturb the existing geometry.  The particle (or string) under study are examples of  test objects.  

% Another reason in reporting on the problem is the popular belief that the nonrelativistic string  can not be couplrd with gravity in the Newton Cartan  framework . This seems to come from the We have written scures of  
 \hskip .5cm   So we see that the analysis of the string by GGT is of highest interest. Note that the Galilean symmetry is inbuilt in the algorithm of GGT. This new theory  is suitable for the purpose  due to the following features:
     
     \begin{enumerate}
     
     \item The algorithm developed in the theory can be used for any model which is symmetric in  flat Galilean spacetime.  The method of approach automatically carries the symmetry along with it. Thus failure to reproduce the flat theory in the appropriate limit as reported in some studies , has no place in this approach.

     \item  The whole calculations are done by a set of rules, derived once for all . There is no fine adjustments during the calculation of a particular problem.
     
     \item The spacetime emerging from our analysis is the Newton- Cartan spacetime. So far we have considered examples from field theory, particle models. {\bf{In this paper we will provide our results for bosonic string model. }}
 \end {enumerate}

 \hskip .5cm It is now time to discuss the organization of the paper.       After introductory section we give a brief review of nonrelativistic particle problem in curved background and in the following section nonrelativistic string in flat spacetime is discussed . If one follows this review of particle theory in section 2 , then the different steps of the gauging of symmetry approach will be clear. Applying the same algorithm to a string action given in section 3  will presumably  give the string action in nonrelativistic curved spacetime. Considering the situation it will be very welcome if the string action thus derived is generally covariant under the N-C transformations. This covariance is explicitly demonstrated in section 5 . In section 6 the geometrical connection is elaborately investigated . Then to ascertain our statements we have introduced a detail canonical analysis which  reflects in the motion of the system in the phase-space. The gauge generator $G$ has been constructed . The gauge transformation generated by $G$ are shown to be equivalent with the diffeomorphism invariances. This completes the invariance issues. In the next section a comparison with earlier results is given . The concluding remarks are contain in section 9.

 \section{Action for Nonrelativistic Particle in Curved Background}

 We start with a short review of the calculations and results for the nonrelativistic particle model treated by the Galilean gauge theory (GGT) \cite{BMN1}. In $3$ dimensional Euclidean space and absolute time the  parametrized action for a nonrelativistic  particle  is given by,
   
\begin{equation}
S = 
\int
\dfrac{1}{2}m\dfrac{ \dfrac{dX^{a}}{d\lambda}\dfrac{dX^{a}}{d\lambda}}{\bigg(\dfrac{dX^{0}}{d\lambda}\bigg)} ~d\lambda
\label{actionum}
\end{equation}
where $a$ denotes  space index and  $\lambda$ is some parameter which changes monotonically and continuously along the world line. A very important aspect of the action (\ref{actionum}) is its invariance under the  reparametrization transform $\lambda \to \lambda^{\prime}(\lambda)$ which is evident from (\ref{actionum}) .

 Often we will use the proper time ($\tau$) as the parameter. The  parameters which are connected with $\tau$ as in the form
 \begin{eqnarray}
 \lambda = A\tau +B
 \end{eqnarray}
are called the affine parameters. Note that  $A$ and $B$ are constant. Henceforth  we will mean the parameter as affine parameter , unless otherwise stated.
\vskip .5cm In our description  the coordinates $X^\rho$ are functions of $\lambda$ , $X^\rho=X^\rho(\lambda)$. One may consider $X^\rho$ as a set of four scalars with respect to the metric on the world line . In case of the particles this is trivial but for strings or higher order branes this is not so as we will see in this paper.

%\begin{figure}{h}

    % \centering
%\begin{tikzpicture}
%\draw (0,0) -- (4,0) -- (4,4) -- (0,4) -- (0,0);
%\draw (0,0) .. controls (0,4) and (4,0) .. (4,4);
%Tiksline.png 
%label{plot}
%\draw ( 2
%\end{tikzpicture} 
 %\caption{Fig 1: the space time diagram of a relativistic particle}
 
 % \end{figure}

 The action (\ref{actionum}) can be reduced to the usual form of the action by considering the motion in space with absolute time ,
 
\begin{equation}
S = 
\int
\dfrac{1}{2}m \dfrac{dX^{a}}{dX^{0}}\dfrac{dX^{a}}{d X^{0}} dX^{0}
\end{equation}
 
now as $X^{0}=t=t(\lambda)$ and $X^{a}=X^{a}(\lambda)$ , thus 
\begin{equation}
S = 
\int
\dfrac{1}{2}m \dfrac{dX^{a}}{dt}\dfrac{dX^{a}}{dt} dt = \int
\dfrac{1}{2}m (v^{a})^{2} dt
\end{equation}
 It is not difficult  to show that the action (\ref{actionum}) invariant under the global Galilean transformations,
 \begin{equation}
x^\rho \to x^\rho + \xi^\rho ; \xi^0 =  -\epsilon, \xi^k = \eta^k - v^k t; \eta^k= \omega^k{}_l x^l+ \epsilon^k\label{globalgalilean} 
\end{equation}

Note carefully that the symbol $X$ is playing a dual role , on the one hand it is  a coordinate transforming under spacetime transformation (\ref{globalgalilean}) as

\begin{equation}
X^\rho \to X^\rho + \xi^\rho ; \xi^0 =  -\epsilon, \xi^k = \eta^k - v^k t; \eta^k= \omega^k{}_l X^l+ \epsilon^k \label{globalgalilean1} 
\end{equation} 

On the other hand they are dynamical variables governed  by the action (\ref{actionum}). Under the transformation (\ref{globalgalilean1}) ,
 
\begin{equation}
\delta \dfrac{dX^{0}}{d\lambda} = \dfrac{d}{d\lambda}(\delta X^{0}) = - \dfrac{d\epsilon}{d\lambda} = 0
\label{der21}
\end{equation}
%\vspace*{25mm}
as $\epsilon $ is constant and,

\begin{equation}
\delta \dfrac{dX^{k}}{d\lambda} = w^{k}{}_{j}\dfrac{dX^{j}}{d\lambda} - v^{k}\dfrac{dX^{0}}{d\lambda}
\label{der2x}
\end{equation} 
 Hence the transformation of Lagrangian is given by 
 \begin{equation}
\delta L = -\dfrac{d}{d\lambda}\bigg(mv^{k}\dfrac{dX^{k}}{d\lambda}\bigg)
\label{variation1}
\end{equation}
  Since the variations of the fields at the boundary vanishes , therefore the action remains the same.

\vskip .25cm  According to GGT , to couple the theory with gravity we have to replace the ordinary derivatives  $\frac{dX^{\alpha}}{d\lambda}$ by the covariant derivatives  $\frac{DX^{\alpha}}{d\lambda}$\cite{BMN1}, where
\begin{equation}
\dfrac{DX^{\alpha}}{d\lambda} = \dfrac{dX^{\rho}}{d\lambda} \Lambda_{\rho}{}^{\beta}\partial_{\beta}X^{\alpha} = \dfrac{dX^{\rho}}{d\lambda} \Lambda_{\rho}{}^{\alpha}
\label{red1}
\end{equation}
Here $\Lambda_{\rho}{}^{\alpha}$ are a set of new gauge fields. So the action (\ref{actionum}) becomes

\begin{equation}
S = 
\int
\dfrac{1}{2}m\dfrac{ \dfrac{DX^{a}}{d\lambda}\dfrac{DX^{a}}{d\lambda}}{\bigg(\dfrac{DX^{0}}{d\lambda}\bigg)} ~d\lambda
\label{actionm2}
\end{equation}

It is to be noted that   in flat limit the  covariant derivatives $\frac{DX^\alpha}{d\lambda}$ must  go to the ordinary derivatives $\frac{d X^\alpha}{d\lambda}$ , such that  the modified  curved spacetime theory (\ref{actionm2}) smoothly transforms to the original flat spacetime theory   (\ref{actionum}) .  This gives the following condition 

\begin {equation}
\Lambda_\rho{}^\alpha \longrightarrow \delta^\alpha _\rho
\label{NS}
\end {equation}
The condition (\ref{NS}) shows that $ \Lambda_\rho{}^\alpha $  is non-singular. Thus the corresponding matrix is invertible. We denoted  $\Sigma_\alpha{}^\sigma $ as the inverse of ${\Lambda_\rho}^{\alpha}$ .  In the geometric interpretation of our theory we  see that this observation is instrumental.  The transformations of this new gauge fields  $ \Lambda_\rho{}^\alpha $ is such that under local Galilean transformations the  covariant derivatives transform  in the same way as the usual derivatives do under the global Galilean transformations. Thus the  transformation of the covariant derivatives,

\begin{equation}
\delta \dfrac{DX^{0}}{d\lambda} = 0
\label{covder12}
\end{equation}
and,

\begin{equation}
\delta \dfrac{DX^{k}}{d\lambda} = w^{k}{}_{j}\dfrac{DX^{j}}{d\lambda} - v^{k}\dfrac{DX^{0}}{d\lambda}
\label{covder2n}
\end{equation}

Using these two relations we get the transformations of the newly introduced fields, which are given by \cite{BMM2}\cite{BMN1},

\begin{eqnarray}
\delta \Lambda_0{}^0 &=& \dot\epsilon \Lambda_0{}^0 \nonumber\\
\delta \Lambda_i{}^a &=& \omega^a{}_b \Lambda_i{}^b - \partial_i\xi^k \Lambda_k{}^a \nonumber\\
\delta \Lambda_0{}^a &=& \dot\epsilon \Lambda_0{}^a - v^a\Lambda_0{}^0 - \partial_0 \xi^k \Lambda_k{}^a + \omega^a{}_b \Lambda_0{}^b
\label{old1x}
\end{eqnarray} 
while the other field $\Lambda_i{}^0$ simply vanishes.

 Exactly the same procedure will be applied   for the string, as 
the latter is viewed here as an extension of the nonrelativistic particle model in \cite{BMN1}, where it was derived first in this approach. But unlike the particle,  string is an extended object . So the motion of the string sweeps a two dimensional subspace called the world sheet of the string . Now the string   is a relativistic object so the world sheet metric will be relativistic. Hence by the nonrelativistic string we mean such low energy processes where the relativistic excitation do not appear. Thus in case of the string we have to differentiate the phenomena taking place parallel to the world sheet from transverse direction of the bulk.  The transverse motion is only excited in case of nonrelativistic  string.

    The review given above is instructive and it will be a gratifying news that this simple procedure is sufficient for obtaining a field theory in the curved background as has been demonstrated in numerous examples workout in the last decade \cite{BMN1}-\cite{Bn}.  This will be further illustrated in the following section for coupling a nonrelativistic string model  with background gravity.   

 \section{ Nonrelativistic Nambu - Goto  action for the bosonic string }
   The  string is an extension of the particle model as discuss above. Both are relativistic objects  and a particular type of nonrelativistic limit is to be taken to obtain the coupling of the nonrelativistic theories with gravity.   
However the string is an extended object . So the motion of the string traces a two dimensional  world sheet .  This world sheet is mapped by two coordinates, $\sigma$ and   $\tau$, where $\sigma$ is space-like  and  $\tau$ is time-like . For the bosonic string the relativistic Nambu Goto action is given by,
\begin{equation}
S_{\rm{NG}} =   -N\int{d\sigma}{d\tau} \sqrt{- \det h_{ij}}\label{10000}
\end{equation}
where $h_{ij }$ is the metric induced by the target space and given by,
\begin{equation}
h_{ij} = \eta_{\rho\lambda}\partial_i X^\rho \partial_j  X^\lambda 
\end{equation}
where $X^{\rho}=X^{\rho}(\tau,\sigma)$ and $\eta_{\rho\lambda} = {\rm{diag}1,\quad -1, \quad -1 ....}$ is the Lorentzian metric in the target space. Let the string is embedded in a (D+1) dimension and at a particulate time it cuts the embedding space along  $X^{1}$ coordinate line. Thus $X^{0}$ and  $X^{1}$ are longitudinal to the string and the others are transverse. It is to be noted that in place of $X^{1}$ we could take any of  $X^{2},................$ $X^{D}$.  From  (\ref{10000}) by taking $c \to \infty $ limit, we get the nonrelativistic Lagrangian \cite{BMMS},

 \begin{eqnarray}
 {\mathcal{L}}_{NG} = -N{\left(2\epsilon_{\mu \nu}\dot{X}^\mu {X^\prime}{}^\nu\right)}^{-1}\left( \dot{X}^\mu X^{a \prime} -  \dot{X}^a X^{\mu \prime}\right)^2
\label{1001}
 \end{eqnarray}

 and the action is given by
  \begin{equation}
  S_{NG} = \int d\sigma d\tau  {\mathcal{L}}_{NG} 
  \label{NGaction1}
  \end{equation}
with  $  {\mathcal{L}}_{NG}$ given by (\ref{1001}) is Galilean invariant. The derivation of (\ref{1001}) is done in detail in \cite{BMMS}. However for clarity a shot review given in the following.

 Expanding the relativistic  Nambu Goto action of the bosonic string (\ref{10000})  we get,

\begin{equation}
 S_{\mathrm{NG}}  = -N\int{d\sigma}{d\tau}\bigg[\bigg(\frac{\partial X^{\rho}}{\partial\tau}\frac{\partial X_{\rho}}{\partial\sigma}\bigg)^2-\bigg(\frac{\partial X^{\rho}}{\partial\tau}\frac{\partial X_{\rho}}{\partial\tau}\bigg)\bigg(\frac{\partial X^{\rho}}{\partial\sigma}\frac{\partial X_{\rho}}{\partial\sigma}\bigg)\bigg]^\frac{1}{2}
 \label{rellag}
 \end{equation}

Where $X^{\rho}=X^{\rho}(\tau,\sigma)$ , describes the coordinates of a point on the string
 and $\rho$ stands for  the coordinates of the background .

We chose  $c \to \infty $ limiting procedure , among the different approaches that are possible for taking the nonrelativistic limit \cite{BGM1} . As in low energy scenario the slope of the transverse vibration of string is very small , so $\dot{X^1}<<c$ and $d{X^a}<<d{X^1}$. Thus the nonrelativistic limit of the action (\ref{rellag}) is

%\end{document}
\begin{eqnarray}
 S_{\mathrm{NG}}  &=&   -N\int{d\sigma}{d\tau}\left(c\left( \dot{t}X^{\prime 1}-\dot{X^1}t^{\prime}
 \right)\right)\left[1+\frac{\sum_{a}c^2\left(\dot{t}X^{\prime a}-\dot{X^a}t^{\prime}\right)^2}
 {c^2\left(\dot{t}X^{\prime 1}-\dot{X^1}t^{\prime}\right)^2}\right.\nonumber\\
 &-&\left.\frac{\sum_{a}\left(\dot{X^1}X^{\prime a}
 - \dot{X}^a X^{\prime 1}\right)^2} {c^2\left(\dot{t}X^{\prime 1}-\dot{X}^1 t^{\prime}\right)^2}
  \right]^{\frac{1}{2}}
\end{eqnarray}

where   prime over a symbol implies differentiation with respect to $\sigma$ and  dot as a superscript  denotes derivative with respect to $\tau$. Now  taking upto second order term of the small quantities (also making sum over k implicit) and dropped the boundary value term , we get

\begin{eqnarray}
 {\mathcal{L}}_{NG} =-N\bigg[\frac{\bigg(\dot{X^0}X^{\prime a}-\dot{X^a}X^{\prime 0}\bigg)^2}{2\bigg(\dot{X^0}X^{\prime 1}-\dot{X^1}X^{\prime 0}\bigg)}-\frac{\bigg(\dot{X^1}X^{\prime a}-\dot{X^a}X^{\prime 1}\bigg)^2}{2\bigg(\dot{X^0}X^{\prime 1}-\dot{X^1}X^{\prime 0}\bigg)}\bigg]\label{1000a}
\end{eqnarray}
Here $X^{0}=ct$. This equation (\ref{1000a}) is the nonrelativistic  Lagrangian  for  bosonic string . Using covariant notation we get (\ref{1001}) from (\ref{1000a}).

 The nonrelativistic Lagrangian (\ref{1001}) can also be written as

 \begin{eqnarray}
 {\mathcal{L}}_{NG} = -N{\left(\epsilon_{\mu \nu}\sigma_{\alpha\beta} \frac{\partial X^{\mu}}{\partial\sigma_{\alpha}}\frac{\partial X^{\nu}}{\partial\sigma_{\beta}}\right)}^{-1}\left( \epsilon_{\alpha \beta}\frac{\partial X^{\mu}}{\partial\sigma_{\alpha}}\frac{\partial X^{a}}{\partial\sigma_{\beta}}\right)^2
\label{1002}
 \end{eqnarray}

 Where $(\mu,\nu) = 0,1$ ,  $ a = 2........ D $ , $(\alpha,\beta) = 1,2$  ,  and $\sigma_{1} = \tau$ and $\sigma_{2} = \sigma$.

 Our Lagrangian (\ref{1002}) is invariant under the  Galilean transformations \cite{BMMS} ,
 
 \begin{eqnarray}
 \delta X^0 &=&    -\epsilon \nonumber\\
 \delta X^1 &=& \epsilon^1   - v^1 X^0\nonumber\\
 \delta X^a &=& \epsilon^a +  {\omega^a}_lX^l - v^a X^0 ; k,l > 1
 \label{gt}
\end{eqnarray}

 With this we finish the review of the method applied to a 
 generally covariant particle model and introduce the main results of our construction of the string (bosonic) action
 following the particle action, we are in a position to introduce gravity. For convenience and clarity we take the string action in the Nambu-Goto form in the following section.
 
 \section{ Nonrelativistic bosonic string in curved background from Galilean gauge theory}
 
 Coupling a Nonrelativistic string theory with gravity has been found to be a difficult task  because of the peculiar geometry of the nonrelativistic string embedded in curved spacetime. The string is inherently a relativistic object. So the 2 dimensional slice of the spacetime produced by the motion of the string (the world sheet) has a Minkowski  metric  structure where as the transverse bulk has an Euclidean structure. The point is that the nonrelativistic description for the string is relevant for low energy excitations, thus in interaction with gravity the world sheet is not affected  \cite{p}. Remember  that in formulating the nonrelativistic action for flat spacetime we have enforced the condition $\omega^{1a}=0$ , where $\omega$ is the spatial rotation parameter , the coordinate axis $X^{1}$ is longitudinal and $X^{a}$
 are transverse to world sheet .  
  
                       Once the peculiarity of the nonrelativistic string geometry is understood it is simple to write down the corresponding action for such string coupled with curved manifold , thanks to the algorithm of Galilean gauge theory (GGT) . We will thus replace the ordinary derivative by covariant derivative \cite{BMN1}. Note that this replacement is  with in the transverse part of the manifold. 
   
    Hence the Lagrangian of nonrelativistic bosonic string (\ref{1002}) in curved background is given by

    \begin{eqnarray}
 {\mathcal{L}}_{NG} = -N{\left(\epsilon_{\mu \nu}\sigma_{\alpha\beta} \frac{\partial X^{\mu}}{\partial\sigma_{\alpha}}\frac{\partial X^{\nu}}{\partial\sigma_{\beta}}\right)}^{-1}\left( \epsilon_{\alpha \beta}\frac{\partial X^{\mu}}{\partial\sigma_{\alpha}}\frac{D X^{a}}{d\sigma_{\beta}}\right)^2
\label{1007}
 \end{eqnarray}
  Which is a direct generalization of the nonrelativistic  particle model \cite{BMN1}. Explicitly 

\begin{eqnarray}
\dfrac{DX^{a}}{d\sigma_{\beta}} &=& \dfrac{\partial X^{l}}{\partial \sigma_{\beta}} \Lambda_{l}{}^{a}
\nonumber \\
%\end{eqnarray}
%\begin{equation}
\delta \dfrac{DX^{a}}{d\sigma_{\beta}} & = &\dfrac{D X^{b}}{d\sigma_{\beta}} \omega^a{}_b - \dfrac{\partial X^{0}}{\partial\sigma_{\beta}} u^{a}
\nonumber \\
%\end{equation}
%\begin{equation}
\delta \dfrac{\partial X^{\mu}}{\partial\sigma_{\alpha}} & = &\dfrac{\partial X^{\nu}}{\partial \sigma_{\alpha}} \omega^\mu{}_\nu 
\label{n1}
\end{eqnarray}

 Where $(\mu,\nu) = 0,1$  and  $ (a,b,l,m) = 2........ D $ . Note that those degrees of freedom transverse to the string world sheet is interacted by the nonrelativistic
gravity. We have already discussed this issue in the above. Using these relations (\ref{n1}) we get the transformations of the newly introduced fields, which are given by

\begin{eqnarray}
\delta \Lambda_l{}^a &=& \omega^a{}_b \Lambda_l{}^b - \partial_l\xi^m \Lambda_m{}^a 
\label{1old1x}
\end{eqnarray} 
while the variation of remaining fields  simply vanishes.

\vskip.5cm In the above we have seen that the construction of string action in curved spacetime is just the repetition  of what we have done in the particle model. The theory  (\ref{1007}) is generally covariant according to the premises of GGT but one may still be sceptic about the validity of GGT \footnote{ Upto the present time GGT have been applied to scores of problem with flying colors erasing what was confusing in the field . }. So we  would like to explicitly check the covariance of  (\ref{1007}) under the general coordinate transformation. The corresponding results are given in the following section.

\section{General covariance of the theory}
  
Our next task is to show that the  (\ref{1007}) is invariant under local Galilean transformation. This is equivalent to that
the diffeomorphism invariance
\begin{equation}
X^{\rho} \to X^{\rho} + \xi^{\rho}\label{1008}
\end{equation}
where 
\begin{eqnarray}
\xi^{\mu}  &=&\epsilon^{\mu} + \omega^\mu{}_\nu  X^{\nu}  ~~,~~  \mu,\nu = 0,1   \nonumber\\        \xi^{a} &=& \epsilon^{a} + \omega^a{}_l X^{l} - u^{a} t  ~~,~~ a = 2........ D .
  \end{eqnarray} 
holds.
  
  Since degrees of freedom   transverse to the string world sheet is interacted by the nonrelativistic
gravity , thus the transformation parameters $\epsilon^{a}$,$ u^{a}$ and $\omega^a{}_l $ are arbitrary functions of coordinates and time .  The variation of Lagrangian (\ref{1007}) due to the spacetime diffeomorphism is ,

    \begin{eqnarray}
 \delta{\mathcal{L}}_{NG} &=& -N{\left(\epsilon_{\mu \nu}\sigma_{\alpha\beta} \frac{\partial X^{\mu}}{\partial\sigma_{\alpha}}\frac{\partial X^{\nu}}{\partial\sigma_{\beta}}\right)}^{-1}\bigg[\delta\left( \epsilon_{\alpha \beta}\frac{\partial X^{\mu}}{\partial\sigma_{\alpha}}\frac{D X^{a}}{d\sigma_{\beta}}\right)^2\bigg] \nonumber\\          &  &\qquad \qquad  -N\bigg[\delta{\left(\epsilon_{\mu \nu}\sigma_{\alpha\beta} \frac{\partial X^{\mu}}{\partial\sigma_{\alpha}}\frac{\partial X^{\nu}}{\partial\sigma_{\beta}}\right)}^{-1}\bigg]\left( \epsilon_{\alpha \beta}\frac{\partial X^{\mu}}{\partial\sigma_{\alpha}}\frac{D X^{a}}{d\sigma_{\beta}}\right)^2 \label{1009}
 \end{eqnarray}  
  
 Now the calculation of (\ref{1009}) is very simple but quite lengthy , so we calculate it in term by term.  The 2nd term of the (\ref{1009})

 \begin{eqnarray}
 \delta{\left(\epsilon_{\mu \nu}\sigma_{\alpha\beta} \frac{\partial X^{\mu}}{\partial\sigma_{\alpha}}\frac{\partial X^{\nu}}{\partial\sigma_{\beta}}\right)}^{-1} & =&  - \frac{ \epsilon_{\mu \nu}\sigma_{\alpha\beta} \bigg[\omega^\mu{}_\xi \frac{\partial X^{\xi}}{\partial\sigma_{\alpha}}\frac{\partial X^{\nu}}{\partial\sigma_{\beta}} + \omega^\nu{}_\xi \frac{\partial X^{\mu}}{\partial\sigma_{\alpha}}\frac{\partial X^{\xi}}{\partial\sigma_{\beta}}\bigg]}{\bigg(\epsilon_{\mu \nu}\sigma_{\alpha\beta} \frac{\partial X^{\mu}}{\partial\sigma_{\alpha}}\frac{\partial X^{\nu}}{\partial\sigma_{\beta}}\bigg)^2}      \nonumber
 \end{eqnarray}

 as $\epsilon_{\mu \nu}$ is anti symmetric and $\sigma_{\alpha\beta}$ is symmetric thus 
  \begin{eqnarray}
  \delta{\left(\epsilon_{\mu \nu}\sigma_{\alpha\beta} \frac{\partial X^{\mu}}{\partial\sigma_{\alpha}}\frac{\partial X^{\nu}}{\partial\sigma_{\beta}}\right)}^{-1} = 0 \label{1010}
\end{eqnarray}

now putting the value of $\dfrac{DX^{b}}{d\sigma_{\beta}}$ in 1st term of the (\ref{1009}) we get,

\begin{eqnarray}
\delta\left( \epsilon_{\alpha \beta}\frac{\partial X^{\mu}}{\partial\sigma_{\alpha}}\frac{D X^{a}}{d\sigma_{\beta}}\right)^2 &=&\bigg( 2\epsilon_{\lambda \xi}\frac{\partial X_{\mu}}{\partial\sigma_{\lambda}}\frac{\partial X^{m}}{\partial\sigma_{\xi}}\Lambda_{m}{}^{a}\bigg)\bigg[\epsilon_{\alpha \beta}\omega^{\mu}{}_{\nu}\frac{\partial X^{\nu}}{\partial\sigma_{\alpha}}\frac{\partial X^{l}}{\partial\sigma_{\beta}}\Lambda_{l}{}^{a} + \nonumber\\          &  &\qquad \qquad   \bigg(\omega^{a}{}_{b}\dfrac{\partial X^{l}}{\partial \sigma_{\beta}} \Lambda_{l}{}^{b} - u^{a}\dfrac{\partial X^{0}}{\partial\sigma_{\beta}} \bigg)\frac{\partial X^{\mu}}{\partial\sigma_{\alpha}}\epsilon_{\alpha \beta}\bigg] \label{1011}
\end{eqnarray}

the 3rd term of  the equation  (\ref{1011})  

\begin{eqnarray}
2 u^{a}\epsilon_{\lambda \xi}\frac{\partial X_{\mu}}{\partial\sigma_{\lambda}}\frac{\partial X^{m}}{\partial\sigma_{\xi}}\Lambda_{m}{}^{a} \bigg(\epsilon_{\alpha \beta}\dfrac{\partial X^{0}}{\partial\sigma_{\beta}}\frac{\partial X^{\mu}}{\partial\sigma_{\alpha}}\bigg)&=&  2 u^{a}\epsilon_{\lambda \xi}\Lambda_{m}{}^{a}\frac{\partial X^{m}}{\partial\sigma_{\xi}}\bigg[\frac{\partial X_{0}}{\partial\sigma_{\lambda}} \bigg(\dfrac{\partial X^{0}}{\partial\sigma_{1}}\frac{\partial X^{0}}{\partial\sigma_{2}}-\nonumber\\          &  &\qquad \qquad \dfrac{\partial X^{0}}{\partial\sigma_{2}}\frac{\partial X^{0}}{\partial\sigma_{1}}\bigg)+\frac{\partial X_{1}}{\partial\sigma_{\lambda}} \bigg(\dfrac{\partial X^{0}}{\partial\sigma_{1}}\frac{\partial X^{1}}{\partial\sigma_{2}}-\nonumber\\          &  &\qquad \qquad \dfrac{\partial X^{0}}{\partial\sigma_{2}}\frac{\partial X^{1}}{\partial\sigma_{1}}\bigg)\bigg]\label{1012}
\end{eqnarray}
 the 1st term of (\ref{1012})  is zero and if we swap $\sigma_{1}$,$\sigma_{2}$ then 2nd term also becomes zero. Swapping indices and using the different symmetry property we find  the first term and second terms of (\ref{1011}) vanish. 
 
\begin{eqnarray}
\delta\left( \epsilon_{\alpha \beta}\frac{\partial X^{\mu}}{\partial\sigma_{\alpha}}\frac{D X^{a}}{d\sigma_{\beta}}\right)^2 = 0\label{1013}
\end{eqnarray}

So from (\ref{1013}) and (\ref{1010}) we see that

   \begin{eqnarray}
 \delta{\mathcal{L}}_{NG} = 0 \label{786}
 \end{eqnarray}

Thus the curved background Lagrangian  (\ref{1007}) is invariant under diffeomorphism due to local Galilean transformations.
The action (\ref{1007}) is quite satisfactory.
 The flat limit poses no problems. In this limit  the vielbeins reduce to the Kronecker deltas and using (\ref{n1})  we can easily show that the action (\ref{1007}) reduces to the  NR string action (\ref{1002}) .
 
 In this connection one may note that in the examples that are available in the literature \cite{bag1} \cite{bag} , there seems to be a no-go statement for the nonrelativistic string. But there is no contradiction of this with our results as in the models where the no-go statement is applicable , assumed a different geometric structure for NRDI than ours. They include a all pervading  gauge field in the conventional Newton-cartan structure for different reasons which are out of the scope of the present paper. It is suffice to say that we follow the conventional Newton-Cartan geometry \cite{Cartan-1923}\cite{Cartan-1924}.

\section{Geometrical Connection}

The gauging of symmetry is not an unknown process of physics. The whole particle physics in the standard model is defined by gauging of  $U(1)$ X $SU(2)$ X $SU(3)$ group.  Utiyama  use the gauge principle in connection with  relativistic geometry \cite{Uti}. Similarly in nonrelativistic scenario Galilean gauge theory (GGT) leads Newton-Cartan  manifold.  In GGT to account for the local degrees of freedom a local coordinate basis is set up at every point where the basis vectors trivially parallel to the respective global basis vectors \cite{blago}.
 The transformations of the gauge fields appeared during localization have a suggestive expression which are begging for a geometrical interpretation
\cite{BM4} . For instance look at (\ref{1old1x})  the gauge field $\Lambda_l{}^a$ carry two set of indices one referring to the global co-ordinate and the other to the local co-ordinate\footnote{ local basis denoted by indices from beginning of alphabet ($a, b $ etc. or $\alpha$,$\beta$ etc.), where as  global basis denoted by those from the middle ( $i, j$ etc. or $\mu$ ,$\nu$ etc.).}.
If we don't consider the origin of equation (\ref{1old1x}) but look at it as a geometric variation of the vielbeins of a curved manifold , then it is simple to interpret it as a diffeomorphism $\xi^{a}$ of the manifold . This is the  
pinnacle point of the logical built-up of GGT where we have identified the previous gauge field in the new geometric avatar. We will show in the following how the qualitative discussion here can be translated  into quantitative relationship and see that the corresponding metric elements satisfies the Newton-Cartan algebra.

  The gauge fields ${\Lambda_\rho}^{\gamma}$ becomes equal to identity in a continuous manner in the flat limit (see discussion around equation (\ref{NS})) and hence the corresponding  matrix is non-singular. Thus it has a unique inverse.
  
   We introduced  $\Sigma_\gamma{}^\sigma $  as  the inverse of ${\Lambda_\rho}^{\gamma}$  , 
\begin{equation}
\Sigma_\gamma{}^\rho{\Lambda_\rho}^{\epsilon} =\delta^\epsilon_\gamma \hspace{.2cm};\hspace{.2cm}
\Sigma_\gamma{}^\rho{\Lambda_\sigma}^{\gamma} =\delta^\rho_\sigma \label{B}
\end{equation}

From  the transformations of the $\Lambda's$ and using the above relations (\ref{B}) , it is easy to compute the transformations of the $\Sigma's$   \cite{BMM2}. The global and local  basis are connected by these vielbeins in the following way,

\begin{equation}
\hat e_\rho = \Lambda_\rho{}^\gamma\hat e_\gamma, \,\,\, \hat e_\gamma = \Sigma_\gamma{}^\rho \hat e_\rho
\label{basis}
\end{equation}

For flat spacetime, there is no difference between the global and local  basis , as the  vielbeins reduces to Kronecker deltas  in flat spacetime.

 For the string model (\ref{1007}) $\Lambda_{0}{}^{0}=1$, $ \Lambda_1{}^1=1$ , $\Lambda_l{}^a $ are arbitrary and all other $\Lambda's$ simply vanishes. Using (\ref{B}) we can easily  get the values of different  $\Sigma's$ which are $\Sigma_{0}{}^{0}=1$, $ \Sigma_1{}^1=1$  and  $\Sigma_l{}^a$ are arbitrary , while all other $\Sigma's$ simply zero.

It is well known that  the Newton-Cartan geometry is a degenerate manifold with  a one form $\tau_\rho $ and a  singular metric $h^{\rho\sigma}$ satisfying the following algebra,
\begin{eqnarray}
h^{\rho\sigma} \tau_\sigma = 0 \quad\quad  h_{\rho\sigma} \tau^\sigma = 0  \quad\quad  \tau ^\rho \tau_\rho = 1\nonumber\\
h_{\rho\sigma} h^{\sigma\phi} = P^\rho _\phi =\delta^\rho _\phi - \tau^\rho\tau_\phi
\label{algebra}
\end{eqnarray}
where, $P^\rho _\phi$ is the projection  operator.   The quantities $\tau^\rho$ and  $h_{\rho\sigma}$ are additional structures defined for  lowering or raising indices.

In the following we  show that our candidates for the vielbeins and their inverse can be used to build up the metric structures of the manifold  which satisfy the set (\ref{algebra}).
 
  We can define the  metrics as ,
   
 \begin{equation}
h^{\rho\sigma}={\Sigma_i}^{\rho}{\Sigma_i}^{\sigma}; \hspace{.2cm}\tau_{\rho}={\Lambda_\rho}^{0} 
\label{spm}
\end{equation}

and,
\begin{equation}
h_{\rho\sigma}=\Lambda_{\rho}{}^{i} \Lambda_{\sigma}{}^{i}; \hspace{.2cm}\tau^{\rho}=
{\Sigma_0}^{\rho}\hspace{.3cm}
\label{spm2}
\end{equation}

 Where $(\rho,\sigma,\phi,\gamma,\epsilon) = 0,1 ,2........ D $ , $(i,j)=1 ,2........ D$ and $ (a,b,l,m) = 2........ D $ .

Now 
\begin{eqnarray}
h^{\rho\sigma} \tau_\sigma = {\Sigma_i}^{\rho}{\Sigma_i}^{\sigma}{\Lambda_\sigma}^{0} = {\Sigma_i}^{\rho} \bigg({\Sigma_i}^{0}{\Lambda_0}^{0}+{\Sigma_i}^{1}{\Lambda_1}^{0}+{\Sigma_i}^{l}{\Lambda_l}^{0} \bigg)
\end{eqnarray}
 For string model (\ref{1007}) $({\Sigma_i}^{0},{\Lambda_1}^{0},{\Lambda_l}^{0})=0$ . Thus
 
 \begin{eqnarray}
h^{\rho\sigma} \tau_\sigma = 0
\end{eqnarray}
 similarly
\begin{eqnarray} 
  h_{\rho\sigma} \tau^\sigma = 0
\end{eqnarray}

Now as $({\Sigma_0}^{0},{\Lambda_0}^{0})=1$, thus

 \begin{eqnarray}
\tau^{\rho} \tau_{\rho} ={\Sigma_0}^{\rho}{\Lambda_\rho}^{0}=\bigg({\Sigma_0}^{0}{\Lambda_0}^{0}+{\Sigma_0}^{1}{\Lambda_1}^{0}+{\Sigma_0}^{l}{\Lambda_l}^{0}\bigg)=1
\end{eqnarray}

and

\begin{eqnarray}
h^{\rho\phi}h_{\phi\sigma}&=&{\Sigma_i}^{\rho}{\Sigma_i}^{\phi}\Lambda_{\phi}{}^{j} \Lambda_{\sigma}{}^{j}\nonumber\\          
   &        = & {\Sigma_i}^{\rho}\bigg({\Sigma_i}^{0}{\Lambda_0}^{j}+{\Sigma_i}^{1}{\Lambda_1}^{j}+{\Sigma_i}^{l}{\Lambda_l}^{j}\bigg) \Lambda_{\sigma}{}^{j} \nonumber\\ &=& {\Sigma_\alpha}^{\rho}\Lambda_{\sigma}{}^{\alpha}-\tau^{\rho}\tau_{\sigma} 
\end{eqnarray}

using (\ref{B}) we get ,

\begin{eqnarray}
h^{\rho\phi}h_{\phi\sigma}= \delta_{\sigma}^{\rho}-\tau^{\rho}\tau_{\sigma} \label{787}
\end{eqnarray}

\vskip .5cm The above calculations clearly show that the spacetime generated by GGT is the conventional Newton-Cartan manifold. This is further to  note that all these results are obtained from the dynamics of the nonrelativistic string.  We have already noted that other nonrelativistic geometry may be possible. Since there is no possibility of experimental verification in near future, hence one cannot identify any of the variation of NRDI which is the chosen one. The logical consistency is all important. Judging the situation from this angle we can say, in our formulation there is inner consistency of the flow of logic at every stage. This can be further explored by the following calculation. 

\vskip .5cm From (\ref{spm2}) and (\ref{1old1x}) we find,
\begin{equation}
\delta h_{lm} = (\omega^a{}_{b} \Lambda_l{}^b - \partial_l \xi^n \Lambda_n{}^a)\Lambda_m{}^a + (\omega^a{}_{b} \Lambda_m{}^b - \partial_m \xi^n \Lambda_n{}^a)
\Lambda_l^a 
\label{proof11}
\end{equation}

Using anti-symmetric property of  $\omega^a{}_{b}$ and equation (\ref{spm2}) we get ,
\begin{equation}
\delta h_{lm} =  - \partial_l \xi^n h_{nm}  - \partial_m \xi^n h_{nl}
\label{proof1}
\end{equation}

Now it is easy  to express the action (\ref{1007})  using the Newton-Cartan elements ,
\begin{equation}
 S=  -N\int h_{lm}{\left(\epsilon_{\mu \nu}\sigma_{\alpha\beta} \frac{\partial X^{\mu}}{\partial\sigma_{\alpha}}\frac{\partial X^{\nu}}{\partial\sigma_{\beta}}\right)}^{-1}\left( \epsilon_{\alpha \beta}\frac{\partial X^{\mu}}{\partial\sigma_{\alpha}}\frac{\partial X^{l}}{\partial \sigma_{\beta}}\right)\left( \epsilon_{\alpha \beta}\frac{\partial X_{\mu}}{\partial\sigma_{\alpha}}\frac{\partial X^{m}}{\partial \sigma_{\beta}}\right) ~d\sigma d\tau
\label{lagm10}
\end{equation}
 The action (\ref{lagm10})  can be interpreted
as the action of a nonrelativistic   bosonic string coupled with a Newton-Cartan background. 

So far our discussions were based on the Lagrangian formulation . The corresponding Hamiltonian structure is no less important. Accordingly we take up the canonical  analysis in the following section.

 \section{ Canonical analysis }
 
 We have already established our Lagrangian for nonrelativistic string in curved N-C background. Now be followed by   the Hamiltonian analysis. Defining the canonical momenta in the usual way  we get
 
   %{\color{red}{ Now being a reparamtrization invariant theory, it is already covariant \cite{HRT}. Thus the model is  equally interesting example of constrained systems, in comparison with its non-relativistic flat  counterparts. A system with first class constraints  necessarily posseses gauge degrees of freedom \cite{D}. Many properties of such systems can be derived without fixing the gauges. This gauge independent approach has been proved to be very useful \cite{BRR1}\cite{BRR2} . The importance of canonical analysis is well known to us. On the  contrary, this aspect has been less emphasized in the literature. Therefore we will  try to give a holistic account of the topic.}}  Here the fields are $X^0(\tau,\sigma)$ ,$X^1(\tau,\sigma)$ ,$X^k(\tau,\sigma)$ (where k=2,3,....D ). The Lagrangian of non relativistic bosonic string  in curved background is given by

\begin{eqnarray}
\Pi_{0}=\frac{\partial{\mathcal{L}}}{\partial\dot{X^0}} &=& \bigg[X^{\prime l}\Lambda_{l}{}^{a}\bigg(\epsilon_{\mu \nu}\dot{X}^\mu {X^\prime}{}^\nu\bigg)^{-1}(\dot{X}^0 X^{\prime m}\Lambda_{m}{}^{a}-    \dot{X}^m X^{\prime 0}\Lambda_{m}{}^{a}) \nonumber\\          &  &\qquad \qquad    -\frac{X^{\prime 1}}{2} {\bigg(\epsilon_{\mu \nu}\dot{X}^\mu {X^\prime}{}^\nu\bigg)}^{-2}\bigg( \dot{X}^\mu X^{\prime m}\Lambda_{m}{}^{a} -  \dot{X}^m X^{ \prime \mu}\Lambda_{m}{}^{a}\bigg)^2\bigg]
  \label{pi0}
\end{eqnarray}

\begin{eqnarray}
\Pi_{1}=\frac{\partial{\mathcal{L}}}{\partial\dot{X^1}} & =& \bigg[-X^{\prime l}\Lambda_{l}{}^{a}\bigg(\epsilon_{\mu \nu}\dot{X}^\mu {X^\prime}{}^\nu\bigg)^{-1}(\dot{X}^1 X^{\prime m}\Lambda_{m}{}^{a}-    \dot{X}^m X^{\prime 1}\Lambda_{m}{}^{a}) \nonumber\\          &  &\qquad \qquad    +\frac{X^{\prime 0}}{2} {\bigg(\epsilon_{\mu \nu}\dot{X}^\mu {X^\prime}{}^\nu\bigg)}^{-2}\bigg( \dot{X}^\mu X^{\prime m}\Lambda_{m}{}^{a} -  \dot{X}^m X^{ \prime \mu}\Lambda_{m}{}^{a}\bigg)^2\bigg]
  \label{pi1}
\end{eqnarray}

and

 \begin{eqnarray}
 \Pi_{k}=\frac{\partial{\mathcal {L}}}{\partial\dot{X^k}} =\left(-\epsilon_{\mu \nu}\dot{X}^\mu {X^\prime}{}^\nu\right)^{-1}\bigg[\Lambda_{k}{}^{a} X^{\prime}_{\mu}(\dot{X^\mu} X^{\prime m}\Lambda_{m}{}^{a}-\dot{X}^m  X^{\prime\mu}\Lambda_{m}{}^{a} )\bigg]
\end{eqnarray}

 where $\Pi_{0},\Pi_{1}$ and $\Pi_{k}$ are canonical momenta respectively conjugate to the fields  $X^0(\tau,\sigma)$ ,$X^1(\tau,\sigma)$ and $X^k(\tau,\sigma)$ (where k=2,3,....D ). These definition immediately give following primary constraints ,

\begin{eqnarray}
\Omega_{1} &=& \Pi^\rho{ X^\prime}_ \rho \approx 0 \nonumber\\
\Omega_{2}& =&{\bf \Pi}_{k} {\bf \Pi}_{l}\Sigma_{a}{}^{k}\Sigma_{a}{}^{l}  +{\bf {X}^{\prime}}^k {\bf {X}^{\prime}}^l \Lambda_{k}{}^{a}\Lambda_{l}{}^{a}-2\sigma^{\alpha}{}_{\beta}\Pi_\alpha {X^\prime}^\beta  \approx 0
\label{115}
\end{eqnarray}

Where $\sigma^{\alpha}{}_{\beta}$ stands for second Pauli matrix. For this theory the fundamental Poisson's brackets  are given by ,
\begin{eqnarray}
 \{X^{\rho}\left(\tau,\sigma\right),
 \Pi_{\phi}\left(\tau,\sigma^{\prime}\right)\} = \eta_{\phi}^{\rho}
 \delta\left(\sigma - \sigma^{\prime}\right)
\label{116}
\end{eqnarray}
Using (\ref{116}) we can  work out the algebra of the constraints as, 
\begin{eqnarray}
\left\{ \Omega_{1}\left(\sigma\right),\Omega_{2}\left(\sigma^{\prime}\right)
\right\} &=& \left(\Omega_2(\sigma ) +\Omega_2(\sigma^\prime )\right)\delta\left(\sigma -\sigma^\prime \right)\nonumber\\
\left\{ \Omega_{1}\left(\sigma\right),\Omega_{1}\left(\sigma^{\prime}\right)
\right\} &=& \left(\Omega_1(\sigma ) +\Omega_1(\sigma^\prime )\right)\delta\left(\sigma -\sigma^\prime \right)\nonumber\\
\left\{ \Omega_{2}\left(\sigma\right),\Omega_{2}\left(\sigma^{\prime}\right)
\right\} &=& \left(\Omega_1(\sigma ) +\Omega_1(\sigma^\prime )\right)\delta\left(\sigma -\sigma^\prime \right)
\label{constalng}  
\end{eqnarray}

 So  the Poisson brackets between the constraints (\ref{115}) are proved to be weakly involutive . Now let us compute the canonical Hamiltonian, starting from the definition we get,
 
 \begin{eqnarray}
H_{c}(\tau)=\int d\sigma \bigg(\Pi_{\rho}\dot{X^{\rho}}-{\mathcal{L}}\bigg)\label{can}
\end{eqnarray}

putting the value of $\Pi_{\rho}$ ,$\dot{X^{\rho}}$  in (\ref{can}) we get      ,        

 \begin{eqnarray}
H_{c}(\tau)=0
\end{eqnarray}
 This result is indeed gratifying because the theory (\ref{1007}) is already parameterized \cite{HRT}. As the canonical Hamiltonian is zero , thus the total Hamiltonian is given by
   
  \begin{equation}
H_{T} = \int d\sigma \left( \kappa \Omega_1 + \zeta\Omega_{2}\right)
\label{HTOT}
\end{equation}
where $\kappa$ and $\zeta$ are Lagrange multipliers . Also the fact that the total Hamiltonian is a linear combination of the constraints, so

\begin{eqnarray}
\{\Omega_{1}(\sigma),H_{c}(\tau)\}=\int d\sigma^{\prime} \{\Omega_{1}(\sigma),\left( \kappa \Omega_1(\sigma^{\prime}) + \zeta\Omega_{2}(\sigma^{\prime})\right)\}\approx 0 \nonumber\\
\{\Omega_{2}(\sigma),H_{c}(\tau)\}=\int d\sigma^{\prime} \{\Omega_{2}(\sigma),\left( \kappa \Omega_1(\sigma^{\prime}) + \zeta\Omega_{2}(\sigma^{\prime})\right)\}\approx 0  \label{cannn}
\end{eqnarray}

Thus the constraints algebra is closed. All the constraints in the set (\ref{115})  
 are primary and first class. The implication of this is very important in the  construction of the gauge generator as we will see in the following.
   
\subsection { Construction of  the gauge generator  }

According to the discussion in the last section the most general form of the gauge generator can be written as  

  \begin{eqnarray}
  G (\tau) = \int d\sigma \left( \alpha(\sigma)\Omega_1 + \beta (\sigma)\Omega_2\right) 
  \end{eqnarray}
Where $\alpha({\sigma})$, and $\beta({\sigma})$ are the gauge parameters. From the result of the last section we find that all the gauge transformation of the system may be obtained from varying the set $\alpha ({\sigma})$ and $\beta({\sigma})$ . Now this gauge transformation give the following transformation of the fields

\begin{eqnarray}
\delta_{G} X^0 &=& \left[ X^0, G \right]_{PB}= \alpha(\sigma)X^{0 \prime} - \beta(\sigma)X^{1 \prime}\nonumber\\
\delta_{G} X^1 &=& \left[ X^1, G \right]_{PB}= \alpha(\sigma)X^{1 \prime} - \beta(\sigma)X^{0 \prime}\nonumber\\
\delta_{G} X^k &=& \left[ X^k, G \right]_{PB}=\alpha(\sigma)X^{k\prime} + \beta(\sigma)\Sigma_{a}{}^{k}\Sigma_{a}{}^{l} \Pi_{l}
\label{gv1}
\end{eqnarray}

using the mapping 
\begin{eqnarray}
 \alpha &=& \xi_1 +\kappa\xi_2 \nonumber \\
 \beta &=&  \xi_2 \zeta \label{geo}
\end{eqnarray}  

we get 
\begin{eqnarray}
\delta_{G} X^0 &=&  (\xi_1 +\kappa\xi_2 )X^{0 \prime} - (\xi_2 \zeta )X^{1 \prime}\nonumber\\
\delta_{G} X^1 &=& (\xi_1 +\kappa\xi_2 )X^{1 \prime} - (\xi_2 \zeta )X^{0 \prime}\nonumber\\
\delta_{G} X^k &=& (\xi_1 +\kappa\xi_2 )X^{k\prime} + (\xi_2 \zeta )\Sigma_{a}{}^{k}\Sigma_{a}{}^{l} \Pi_{l}
\label{gv}
\end{eqnarray}

But in the following we show  if we identify $\xi_{1}$ and $\xi_{2}$ with the diffeomorphism parameters than these changes (\ref{gv}) are the same as those due to diffeomorphism \footnote{Note that there are some  exceptions of this statement of equivalence of the gauge and reparametrization parameter,offshell\cite{blago}\cite{BBSP} . However it has been shown \cite{BBS} that this difference is due to a pure gauge transformation.}. Now  due to diffeomorphism $(\tau^\prime =   \tau + \delta \tau , \sigma^\prime = \sigma + \delta \sigma)$   variations are,

\begin{eqnarray}
\delta_{D} X^0 &=& \xi_1 X^{0 \prime} + \xi_2 \dot{X}^{0}\nonumber\\
\delta_{D} X^1 &=& \xi_1 X^{1 \prime} + \xi_2 \dot{X}^{1}\nonumber\\
\delta_{D} X^k &=& \xi_1 X^{k \prime} + \xi_2 \dot{X}^{k}
\label{rv}
\end{eqnarray}

where $\xi_1=\delta \sigma$ and $\xi_2=\delta \tau$ .We see that in the expression of variations (\ref{rv}) ``velocities" appears .
To get the one to one correspondence we have to replace $ \dot{X}^\rho $ in (\ref{rv}). For this purpose we use  $ \dot{X}^\rho = \left[ {X}^\rho,H_T \right ] $ ,where 
$H_T$ is the total Hamiltonian.  This calculation for $\rho=k$ gives

\begin{eqnarray}
\dot{X}^{k}=\kappa X^{k \prime} + \zeta \Sigma_{a}{}^{k}\Sigma_{a}{}^{l} \Pi_{l}
\end{eqnarray}
Putting this in (\ref{rv}) we get , 
 \begin{eqnarray}
 \delta_{D} X^k &=& (\xi_1 +\kappa\xi_2)  X^{k \prime} + \xi_2 \zeta \Sigma_{a}{}^{k}\Sigma_{a}{}^{l} \Pi_{l} 
 \end{eqnarray}

This is identical with the last equation of (\ref{gv}). For any component of $\rho$ the same results will be obtained . Thus the mapping (\ref{geo}) shows the complete equivalence of the diffeomorphism invariances   with the gauge symmetries of the model. The implication of this is far reaching. In particular it proves beyond doubt that the embedding of the nonrelativistic string action (\ref{1007}) in Newton-Cartan spacetime is completely feasible. The calculation shown in this paper buttress the foundation of the GGT.

 \section{Comparison with earlier results } 
 
We have already said that the application of the Galilean gauge theory (GGT) in a string model is yet to be done. In this paper we have demonstrated that the string action in the curved space is given by (\ref{lagm10}). The action (\ref{lagm10}) is quite satisfactory. The flat limit poses no problems. In this limit  the vielbeins reduce to the Kronecker deltas . The action (\ref{lagm10}) reduces to the nonrelativistic action for a free string in flat space (\ref{1002}) . The Geometric elements that appear in the action (\ref{lagm10}) has been used to find the degenerate metrics ($h^{\rho\sigma} ,\tau_{\rho}$) in the spacetime manifold. On the top of it we have shown that from their definitions ( see equations (\ref{spm}) and (\ref{spm2})) , the Newton-Cartan algebra (\ref{algebra}) follows. Hence we conclude that the nonrelativistic string theory can consistently be formulated in the spacetime manifold , satisfying the convention Newton-Cartan algebra.
 
 \hskip .2cm The nonrelativistic string model is being studied for quite a long time. Naturally the coupling with gravity has been dealt by many authors where the NRDI was provided by the gauging of algebra approach. A representative paper is \cite{bag1} . There theory has a fundamental difference from ours study . The geometric structure in their  theory is complete by the set  ($\tau^{a}_{\mu}, e^{a\prime}_{\mu},m^{a}_{\mu} $). Note that  the first two of this have definite geometric meaning . These are the vielbeins of our theory and the connection of these with the metric elements are also know ( as in (\ref{spm}) and (\ref{spm2})).  So our conception about the geometric structure of the Newtonian theory is quite in keeping with that of the old masters like Cartan. 
 
\hskip .2cm  The third component of the set  ($\tau^{a}_{\mu}, e^{a\prime}_{\mu},m^{a}_{\mu} $) is not that obvious. Without commenting about that we just note that in the flat limit this extra piece still persists, which means that there is  an all pervading gauge field even in vacuum. 
 
\hskip .2cm It is thus apparent that there would be some difference between  GGT algorithem and the algorithem followed in \cite{bag1}. Since we have no experimental check  but only logical consistency , hence one cannot identify any of the variation of NRDI which is the chosen one.  Note that  we have already discussed this issue (see below the equation (\ref{786}) and (\ref{787})) .

\section{Conclusion}

   The nonrelativistic string theory is being studied for a long time because of the phenomenological attraction of the low energy excitations of the string \cite{ABPR}. However , the basic Lagrangian and Hamiltonian  structures are  much less stuided in the literature . If we compare with the relativistic string then such studies are abundant \cite{HRT}. Historically we find that many intricacies of the relativistic string were resolved by considering such simple models\cite{HRT}\cite{poi}. This is absent in the nonrelativistic counter part. {\bf{ Motivated by this we have studied a bosonic  string freely moving in nonrelativistic spacetime manifold}}. Using the Galilean gauge theory we have coupled the free nonrelativistic string with gravity. The consistency of the construction of the model was verified in differenet ways . The most important part of this work is the conditions on the first order variables (i.e. the vielbeins) that emerge intrinsically from Galileon Gauge theory (GGT) algorithm.  Thus we did not have to introduce any auxiliary   fields or some sort of things post-facto. {\bf{Interestingly conditions lead to  the Newton-Cartan algebra. Thus the arbitrary spacetime that was initially assumed  turned out to be the Newton-Cartan spacetime.}} The reason for emphasizing this  point is that  some  well known studies \cite{bag1} \cite{bag} in the subject , rule out such possibility earlier. But we have already observed that the nonrelativistic diffeomorphism can be of different types \cite{BM4}. Notably in \cite{bag1} a set of elements that specify the underlying geometry  completely, contains an   arbitrary gauge field  and so the spacetime they generated is different from Newton-Cartan spacetime.

 The theory elaborated in this paper  were  also analysised in the Hamiltonian formalism . Now the nonrelativistic string model is endowed with singularities . So the Hamiltonian analysis is interesting in its own right. We followed the constraints Hamiltonian analysis of the singular theory according  to Dirac \cite{D} and  {\bf{found that the set of constraints include only two first class primary constraints}}. All the gauge transformations were  expressed in terms of these constraints . So the gauge generator contains two arbitrary parameters. Thus the increment  of the canonical variables were computed. {\bf{It was gratifying to observed that under a certain mapping ,  derived in this paper the variations of the canonical variables become the same as those transformations in GGT (see discussion below equation (\ref{gv}) )}}. This was actually expected  as both the set of transformations were the off-shell invariances of the same action (\ref{1007}) . {\bf{The analysis given here can be used to study more complicated example , for instance this procedure can be extended to closed string}} .

\section{Acknowledgments}

Sk. moinuddin would like to thank CSIR India for the fellowship provided to him.(File no: 08/606(0005)/2019-EMR-I)

\end{document}